\documentclass[11pt]{article}
\usepackage{amssymb,amsmath,bm,a4,amsthm,array,xy,longtable,graphicx}
\usepackage{rotating}

\newcommand{\eeq}{\end{equation}}
\newcommand{\beq}{\begin{equation}}
\newcommand{\ba}{\begin{array}}
\newcommand{\ea}{\end{array}}
\newcommand{\bea}{\begin{eqnarray}}
\newcommand{\eea}{\end{eqnarray}}
\newcommand{\vev}[1]{\langle #1\rangle}
\newcommand{\vp}{\varphi}

\newcommand{\preprintno}[1]{\vspace{-2cm}{\normalsize\begin{flushright}#1\end{flushright}}\vspace{1cm}}

\newcommand{\lise}{${^7}$Li}

\title{\preprintno{HD-THEP-08-22} Time variation of fundamental couplings and dynamical dark energy}

\author{Thomas Dent,
	Steffen Stern and 
	Christof Wetterich \vspace{0.2cm}\\ 
	{\it Institut f{\" u}r Theoretische Physik, Universit{\" a}t Heidelberg } \\ 
	{\it 16 Philosophenweg, Heidelberg 69120 GERMANY } 
	}

\date{\today}

\begin{document}

\maketitle

\begin{abstract}\noindent
Scalar field dynamics may give rise to a nonzero cosmological variation of fundamental constants. Within different scenarios based on the unification of gauge couplings, the various claimed observations and bounds may be combined in order to trace or restrict the time history of the couplings and masses. If the scalar field is responsible for a dynamical dark energy or quintessence, cosmological information becomes available for its time evolution. Combining this information with the time variation of couplings, one can determine the interaction strength between the scalar and atoms, which may be observed by tests of the Weak Equivalence Principle. We compute bounds on the present rate of coupling variation from experiments testing the differential accelerations for bodies with equal mass and different composition and compare the sensitivity of various methods. In particular, we discuss two specific models of scalar evolution: crossover quintessence and growing neutrino models.
\end{abstract}

\section{Introduction}
Several positive and null results on the time variation of fundamental constants have been reported. In a previous paper \cite{Part1} we reviewed the relevant observational evidence and bounds, and introduced several distinct scenarios with unification of gauge couplings (GUT) \cite{Marciano}. Each scenario is characterized by specific relations between the variations of different Standard Model (SM) couplings: specifically, all fractional variations in SM couplings were taken to be proportional to the variation of one underlying unified coupling or a fixed linear combination of several unified couplings. The different scenarios have different consequences for the interpretation of any signal of nonzero variation. We also divided the observations into several cosmological ``epochs'' and discussed the consistency between them within each epoch, as well as whether a monotonic time evolution between epochs could fit the observations. Up to this point no particular dynamical mechanism for the coupling variations was considered.

In quantum field theory, a possible time variation of couplings must be associated to the time variation of a field. This field may describe a new ``fundamental particle'' or stand for the expectation value of some composite operator. In this paper we discuss the consequences of the simplest hypothesis, namely that the field is a scalar, such that its time varying expectation value preserves rotation and translation symmetry locally, as well as all gauge symmetries of the Standard Model.

Any observation of coupling variations in the cosmological history from nucleosynthesis up to now would imply that the expectation value of this scalar field changes appreciably during this cosmological epoch. Such a ``late time evolution'' (in contrast to the electroweak or QCD phase transition, GUT-phase transition or even inflation) is characteristic for models with a dynamical dark energy or quintessence. Indeed, the potential and kinetic energy of the scalar ``cosmon'' field would contribute a homogeneous and isotropic energy density in the Universe, and therefore lead to dynamical dark energy \cite{Wetterich88}. The time variation of ``fundamental couplings'' is a generic prediction of such models \cite{Wetterich:1987fk}. In fact, it has to be explained why the relative variation is tiny during a Hubble time, while the simplest models for a scalar cosmon field arising from superstrings or other unified frameworks typically lead to relative variations of the order one. Efficient mechanisms have been proposed for explaining the smallness of variations, for example based on the approach to a fixed point \cite{Wetterich:2002wm,Wetterich:2008sx}. Unfortunately, the strength of such mechanisms is not known, such that the overall strength of the scalar coupling to matter remains theoretically undetermined. Several investigations have explored the consequence of scalar quintessence models for the time variation of couplings, as well as models specifically designed to account for coupling variations \cite{Gasperini:2001pc, Dvali:2001dd, Olive:2001vz, Sandvik:2001rv, Chiba:2001er, Barrow:2002ed, Wetterich:2002ic, WetterichCrossoverQ, Farrar:2003uw, Parkinson:2003kf, Doran:2004ek, Lee:2004vm, Wetterich:1994bg}. 

The cosmon coupling to atoms determines both the outcome of tests of the Weak Equivalence Principle and the time variation of couplings in the most recent cosmological epoch, including the present. For the latter, one also needs information for the rate of change of the expectation value of the scalar field, which can be expressed in terms of cosmological observables, namely the fraction in dark energy, $\Omega_h$, and its equation of state, $w_h$. In consequence, the differential acceleration $\eta$ between two bodies 
of different composition, and therefore different specific ``cosmon charge'', can be related to the present time variation of couplings and cosmological parameters as \cite{Wetterich:2002ic}
\beq
\label{eq:DifferentialAccelerationParam}
 \eta \simeq 3.8 \times 10^{-12} \left( \frac{\dot{\alpha}/\alpha}{10^{-15}{\rm y}^{-1}} \right)^2 
 \frac{ F
}{\Omega_h (1 + w_h)}\, .
\eeq
Here the present dark energy fraction is $\Omega_h \approx 0.73$ and $\dot{\alpha}/\alpha$ is the present relative variation of the fine structure constant. 
The ``unification factor'' $F$ encodes the dependence on the precise unification scenario (defined in Section \ref{sec:unif}) and on the composition of the test bodies. In the notation of \cite{Wetterich:2002ic} we have $F = \Delta R_Z (1+\tilde{Q})$, where $\Delta R_Z = \Delta (Z/(Z+N))$ is the difference in the proton fraction $Z/A$ between the two bodies, which takes a value $\Delta R_Z \lesssim 0.05$ for typical experimental tests of the equivalence principle. 
The present dark energy equation of state $w_h$ is close to $-1$; we take $1 + w_h \lesssim 0.1$. The factor 
$F$ will be calculated for our different unified scenarios in 
Section~\ref{sec:UnificationFactor}: for typical test mass compositions we find $1 \leq F\leq \rm{few}\times 10^2$.
 
Once $F$ is fixed, the relation \eqref{eq:DifferentialAccelerationParam} 
allows for a direct comparison between the sensitivity of measurements of $\eta$ versus the measurements of $\dot{\alpha}/\alpha$ from laboratory experiments, or bounds from recent cosmological history, such as from the Oklo natural reactor or the isotopic composition of meteorites.

Our paper is organized as follows.
After a brief summary of our previous results concerning evolution factors for different unified scenarios, we introduce in Section~\ref{sec:Scalar} a cosmologically evolving scalar field as the source of the dynamics of a possible coupling variation. The scalar has both an effect on large-scale cosmological evolution as a consequence of its homogeneous potential and kinetic energy, and may also produce local gravitational effects, violating the Weak Equivalence Principle (WEP) due to its interactions with matter. In Sec.~\ref{sec:WEP} we use both aspects to set further constraints on a possible time variation of couplings. We determine, for each unified scenario, which observational methods are most sensitive to detect late-time or present-day variations. This may enable us to distinguish between scenarios if there is a non-zero signal. 

In Section~\ref{sec:QuintessenceModels} we determine for each scenario the evolution factors that a smooth time evolution of the scalar must satisfy to be consistent with observations, in particular with a detectable nonzero variation at $z\sim 1\,$--$\,3$. We discuss whether a simple model of monotonically varying ``crossover quintessence'' can produce the required behaviour. Section~\ref{sec:Growing} is devoted to a recently proposed type of model \cite{Amendola_grownu,Wett_grownu} where the scalar field evolution is halted by its coupling to neutrinos whose mass increases. Such models solve the ``coincidence problem'' of dark energy and give rise to an interesting scalar field evolution with damped oscillations at recent epochs. These models need not obey bounds derived for a monotonic time variation, and we use instead global fits over all data for some specific choices of cosmological model. In Section~\ref{sec:Conclude} we summarize and discuss our findings.

\section{Unified scenarios and epochs}
\label{sec:unif}
In \cite{Part1} we defined different GUT scenarios for varying couplings. They are based on the unification of gauge couplings at a scale $M_X$ below the Planck mass $M_{\rm P}$, with a unified coupling $\alpha_X$. We considered also variations of the Higgs v.e.v.\ $\vev{\phi}$ and, for SUSY models, of the superpartner mass scale $\tilde{m}$. Under the hypothesis of proportionality of relative variations of the ``unified couplings'' $\alpha_X$, $M_X/M_{\rm P}$, $\vev{\phi}/M_X$, $\tilde{m}/M_X$ we can write 
\beq \label{eq:unifdefs}
\Delta \ln \frac{M_X}{M_{\rm P}} = d_M l, \quad \Delta \ln \alpha_X = d_X l, \quad \Delta \ln \frac{\vev{\phi}}{M_X} = d_H l, \quad \Delta \ln \frac{\tilde{m}}{M_X} = d_S l\,.
\eeq
The ``evolution factor'' $l(z)$ depends on the redshift $z$ and will be linked to the evolving value of a scalar field. If $\alpha_X$ varies nontrivially we may normalise $l$ via $d_X=1$. In models without supersymmetry we set $\alpha_X = 1/40$ and $d_S \equiv 0$, while for supersymmetric theories we take $\alpha_X = 1/24$. As explained in \cite{Part1}, we neglect variations of Yukawa couplings, thus all masses of the charged fermions in the Standard Model vary proportional to $\vev{\phi}$. 

The different unified scenarios investigated in \cite{Part1} can be characterized by the unification coefficients $d_k$ as follows:
\begin{itemize}
\item Varying $\alpha$ alone
\item {Scenario 1:} Varying gravitational coupling, $d_M=1$, $d_X= d_H= d_S= 0$
\item {Scenario 2:} Varying unified coupling, $d_X=1$, $d_M= d_H= d_S= 0$: we distinguish Scenario 2 (non-SUSY) and 2S (SUSY) if relevant
\item {Scenario 3:} Varying Fermi scale, $d_H=1$, $d_M= d_X= d_S= 0$
\item {Scenario 4:} Varying Fermi scale and SUSY-breaking scale, $d_H=d_S=1$, $d_M= d_X= 0$
\item {Scenario 5:} Varying unified coupling and Fermi scale, $d_X=1$, $d_H=\tilde{\gamma}$, $d_M= d_S= 0$: we considered $\tilde{\gamma}=42$ for illustration
\item {Scenario 6:} Varying unified coupling and Fermi scale, $d_X=1$, $d_H=d_S=\tilde{\gamma}$, $d_M= 0$: we considered $\tilde{\gamma}=70$ and $\tilde{\gamma}=25$.
\end{itemize}
Computing the couplings of the Standard Model of particle physics in terms of the unified couplings, all relevant observables are then found in terms of $l$, $d_M$, $d_X$, $d_H$ and $d_S$ in each scenario. This allowed us to reduce each observational result to a constraint on the variation of the underlying unified parameter, or equivalently on $l(z)$. 

For each unified scenario, there are a number of questions that may be addressed within our framework. First, how many of the claimed signals of non-zero variation may be consistently accomodated, given a monotonic variation and given other observational constraints on variations. 

As already stated, atomic clock experiments and WEP violation put bounds on the rate of time variation at present. These bounds can be extrapolated to the past and compared to observational constraints at previous epochs, given some assumption for the form of time variation. For this comparison we will generally assume a constant time derivative and restrict the extrapolation to small redshifts. One may also consider the case of oscillating (non-monotonic) variation. However, this introduces too much freedom to allow a general analysis. Hence we consider a few specific models with oscillations in Sec.~\ref{sec:Growing}.

Second, we may consider specifically BBN. If a nonzero variation occurs at high redshift, as suggested by spectroscopic determinations of $\alpha$ and/or $\mu$, one can ask whether such variation can improve the fit to observationally determined abundances including \lise. If a nonzero variation at BBN is favoured, what is the range of values for the evolution factor $l_6 = l_{\rm BBN}$, and what is the best fit that can be obtained for the BBN abundances?

Third, what additional signals are expected for a moderate increase in observational sensitivity, particularly in the case of tests that at present give null results? In other words, what method could rule out or confirm any given model most efficiently? In section \ref{present} we consider, in particular, probing current claimed variations by atomic clocks or detecting the corresponding scalar couplings by WEP experiments.

In order to simplify the comparison of different observations we defined several ``epochs'' as follows:
\begin{itemize}
\item Epoch 1: Today until Oklo, $z<0.14$
\item Epoch 2: Cosmological probes at low redshift, $0.2 \le z \le 0.8$ 
\item Epoch 3: Intermediate redshift, $0.8 \le z \le 2.4$ 
\item Epoch 4: High redshift, $2.4 \le z \le 10$ 
\item Epoch 5: CMB, $z \approx 1100$
\item Epoch 6: BBN, $z \approx 10^{10}$.
\end{itemize}
Laboratory bounds on the present-day time variation of fundamental couplings are incorporated into Epoch 1 by simple linear extrapolation back to the time of Oklo, 1.8 billion years ago. Then for each scenario we evaluated a weighted average of the observations in each epoch, resulting in ``evolution factors'' $l_{1}$,~\ldots,~$l_6$. The relevant values of evolution factors $l_i$ are listed in Table~4 of \cite{Part1}. For convenience, we reproduce this table in the Appendix of this paper.

We found that no scenario was able to fit all the nonzero signals or hints of variation that we considered with a monotonic time variation. The possible nonzero variations are the Murphy {\it et al.}\ observations of $\alpha$ variation in 143 quasar absorption systems \cite{Murphy:2003mi}; the Reinhold {\it et al.}\ observations of $\mu\equiv m_p/m_e$ variation from $H_2$ spectra at high redshift \cite{Reinhold:2006zn}, and the potential solution of the \lise\ problem of primordial nucleosynthesis (BBN) via variation of quark masses \cite{DSW07,Coc06,Dmitriev03}. However, we were able to obtain monotonic variation within some scenarios, if we considered that one of these nonzero results was affected by an unknown systematic error that increased the uncertainty. For example Scenario 6 with $\tilde{\gamma}=25$ is consistent with monotonic evolution if the more conservative $\mu$ bound of \cite{Wendt08} is substituted for the value of $\mu$ variation obtained in \cite{Reinhold:2006zn}.

\section{Coupling variations from evolution of a \\cosmological scalar field}
\label{sec:Scalar}
We now consider the dynamics of the degree of freedom that gives rise to the variations of couplings. If we simply wrote down an action with the couplings $\alpha_X$, {\it etc.}, promoted to functions of cosmological time, then almost all the symmetries that have been used to determine the form of underlying theories would be explicitly violated. This would leave well-tested properties like Lorentz invariance, charge conservation and energy-momentum conservation without a coherent theoretical explanation, and would destroy any predictivity since physics would depend on arbitrary, unknown functions of time. We must introduce one or more new propagating degrees of freedom, that allow us to take a limit where the fundamental symmetries are exactly restored if the corresponding fields have static values. This will lead to a closed system of dynamical equations giving, in principle, predictions for the cosmological and local/present-day behaviour of couplings. 

The simplest possibility of a new degree of freedom allowing a limit of local Lorentz invariance is a scalar field, which may be fundamental or effective (composite). If the scalar is a gauge singlet, all gauge symmetries and the associated charges are conserved. In cosmology we may consider a homogeneous field that only depends on time. More precisely, the value of the field is statistically homogeneous on large scales over hypersurfaces on which the matter density is also statistically homogeneous. For a very light field weakly coupled to matter the local perturbations are generally small relative to the cosmological evolution. In other words, the evolution of the scalar field in a cluster of galaxies or on Earth does not decouple from the cosmological evolution (in contrast to the gravitational field), such that its cosmological time evolution is reflected in a universal variation of couplings, both on Earth and in the distant Universe \cite{Wetterich:2002ic, Mota:2003tm, Shaw:2005vf}.

Observable coupling variations are proportional to the interaction strength of the scalar field to the constituents of ordinary matter (atoms). In the limits of either weak interaction or a slow time evolution of the scalar the coupling variations may become tiny. It is tempting to identify the scalar with the cosmon that is the key ingredient for dynamical dark energy or quintessence, but we will first keep the discussion general. In any case cosmology will put additional constraints on the evolution of such a field. 

\subsection{Coupling functions}
The assumption of unification implies that the variation of the unified coupling is related via some coupling function $B_X$ to the 
scalar $\varphi$
\beq
 \alpha_X = B_X(\varphi),
\eeq
realized typically by a gauge-invariant kinetic term for the gauge fields in the GUT-model $\propto B_X(\vp)^{-1}F^{\mu\nu}F_{\mu\nu}$ \cite{HillShafi83,Ellis:1989as}.~\footnote{For the pure electromagnetic field the gauge invariant description of a varying fine structure constant was given in \cite{Jordan3739,Bekenstein:1982eu}.} In a linear approximation, this yields
\beq
 \Delta \ln\alpha_X = \Delta\ln B_X(\vp) \simeq 
\frac{B'_X}{B_X} \Delta \vp = d_X l,
\eeq
where the dimensionless scalar $\vp$ is obtained by dividing out the reduced Planck mass $M_{\rm P}\equiv(8\pi G)^{-1/2}$. Similarly, the 
unification mass $M_X(\vp)$ or the Fermi scale $\vev{\phi}(\vp)$ may depend on $\vp$, with
\beq
 \Delta \ln \frac{M_X}{M_{\rm P}} = \Delta \ln B_M(\vp), \qquad \Delta \ln \frac{\vev{\phi}}{M_{\rm X}} = \Delta \ln B_H(\vp).
\eeq
More generally, one has in the linear approximation for the couplings $G_k$ of the Standard Model of particle physics
\beq \label{DlnGdvphi}
 \Delta \ln G_k = d_k l \simeq \beta_k \Delta \vp,
\eeq
where we define 
\beq
\label{eq:DefBetak}
\beta_k = \frac{\partial}{\partial \vp} \ln B_k .
\eeq
For $\beta_X\neq 0$ we may choose $d_X=1$ and $l=\beta_X \Delta \vp$. (As mentioned in Section \ref{sec:unif}, in scenarios where $\alpha_X$ does not vary we take $d_H=1$, $l=\beta_H\Delta\vp$.) All quantities may depend on redshift or the epoch $n$. For the relation (\ref{eq:unifdefs}) to hold it is sufficient that the ratios $\Delta \ln B_k/\Delta \ln B_X = d_k/d_X$ remain constant. 

In this and the next two sections we investigate the hypothesis of monotonic evolution. This assumes first a monotonic evolution of the scalar field towards today's value $\vp_0$. The presently measured laboratory values of the fundamental `constants' correspond to $\vp = \vp_0$. Secondly, the coupling functions $B_k(\vp)$ are also assumed to be smooth and monotonic. Thus
\beq
	\vp(z) - \vp_0 \equiv \Delta \vp(z)
\eeq
and $l(z)$ are monotonic functions of $z$, which can be taken to be increasing. This means that in the space of values of the SM couplings $G_k$, variations at previous epochs should lie along a curve passing through $G_{k,0}$, at distances which 
increase smoothly with increasing $z$.

Beyond an overall linear approximation where the $\beta_k$ are constants (for example, evaluated at the present value of the scalar $\vp_0$) we can consider 
several epochs through their redshifts $z_n$, the corresponding values of the scalar $\vp(z_n)$, the coupling function $B_X(\vp(z_n))$ and the variations $\Delta \ln G_{k,n}$. In a ``local linear approximation'' we define the coefficients $\tilde{\beta}_{k,n}$ by 
\beq \label{eq:defln}
	\Delta \ln G_{k,n} = d_{k} l_n = \tilde{\beta}_{k,n} \Delta \vp(z_n) \approx \beta_k(z_n) \Delta \vp(z_n).
\eeq
We keep our assumption of proportionality of all $\Delta \ln B_k$, such that $\Delta \ln G_k(z_n) = d_k l(z_n) = d_k l_n$ with constant $d_k$. In terms of the coupling functions $B_k$ the evolution factors are defined as 
\beq
\label{eq:defEvolutionCoefficient2}
l_n = \Delta \ln \alpha_{X,n} = \ln B_X(\vp(z_n)) - \ln B_X(\vp_0) = \tilde{\beta}_X \Delta \vp(z_n)
\eeq
for the scenarios 2,5,6, and for the scenarios 3,4 we take
\beq
l_n = \Delta \ln \vev{\phi}_{,n} = \ln B_H(\vp(z_n)) - \ln B_H(\vp_0) = \tilde{\beta}_H \Delta \vp(z_n).
\eeq

\subsection{Spatial variation and WEP-violating forces} 
\label{wep}
In contrast to the direct observations of time varying couplings, tests of the universality of free fall (weak equivalence principle, WEP) do not determine directly the values of fundamental `constants' or their possible variations. However, given our basic assumption of a slow time variation, driven by a light scalar degree of freedom, the current limits on composition-dependent long range forces put bounds on the scalar couplings to different constituents of matter. In our language, they measure or constrain the coefficients $\tilde{\beta}_k$ at $z=0$.

These constraints then imply bounds on the time variation of `constants', since the rate of change of the scalar is bounded above by the cosmological effects of its kinetic energy. If one is allowed to adjust the scalar coupling functions arbitrarily, the WEP force bounds on variations might be evaded, for example through some kind of cancellation. But in general, and for the unified scenarios we discuss, WEP violation places significant bounds on the present-day values of scalar couplings. We will follow the treatment of \cite{Wetterich:2002ic,Dent06} (see also \cite{DamourP}) and consider the couplings of a 
scalar field to matter in the ``Einstein frame'', assuming that they produce a small correction to the usual gravitational acceleration. 

We first briefly discuss the relation between a homogeneous cosmological evolution of the scalar at large scales and local variations at small scales. The equation of motion of the scalar $\Phi(t,\vec{x}) = M_{\rm P} \vp(t, \vec{x})$ in a FRW cosmological background reads
\beq
 \Box \Phi + 3 H \dot{\Phi} = -V'(\Phi) - \frac{m'_\Psi(\Phi)}{m_{\Psi}}
 \rho_\Psi(t,\vec{x}),
\eeq
where $H$ is the Hubble rate $\dot{a}/a$ and $\Psi$ denotes some species of matter with density $\rho_{\Psi}$ and mass $m_\Psi(\Phi)$. The homogeneous solution $\bar{\Phi}(t)$ is given by solving
\beq
 \ddot{\bar{\Phi}} + 3 H \dot{\bar{\Phi}} = -V'(\bar{\Phi}) - 
 \frac{m'_\Psi(\bar{\Phi})}{m_{\Psi}}  \bar{\rho}_\Psi(t),
\eeq
where $\rho_\Psi(t,\vec{x}) = \bar{\rho}_\Psi(t) + \delta \rho_\Psi(t,\vec{x})$ and $\bar{\rho}_{\Psi}$ is the averaged matter density. On subtracting this from the general equation of motion, expanding about $\bar{\Phi}$ and making the approximations that $V''(\bar{\Phi})$ is negligibly small ({\it i.e.}\ the scalar is very light), that the cosmological evolution is very slow, and that the coupling $m'_\Psi/m_\Psi$ is much smaller than $1/M_{\rm P}$, we recover a Poisson-like equation 
\beq
 \nabla^2 (\Phi-\bar{\Phi}) \simeq \frac{m'_\Psi(\bar{\Phi})}{m_{\Psi}} \delta \rho_\Psi,
\eeq
applicable to small regions of space (for example the Solar System) over short periods of time. A more sophisticated treatment \cite{Wetterich:2002ic, Mota:2003tm, Shaw:2005vf} yields similar results in that a local variation sourced by inhomogeneities of matter is superimposed on the slow homogeneous cosmological variation. See also \cite{Barrow:2005sv} concerning bounds on spatial variation in theories with special choices of the scalar couplings.

In consequence, Newton's law for the gravitational attraction between two $\Psi$ particles gets modified by an additional piece which involves the dimensionless cosmon charge $Q_{\Psi}$ \cite{Wetterich:2002ic}
\beq
 V_{N} = - \frac{G m_{\Psi}^2}{r} ( 1 + \alpha_{\Psi}) \; ,
\eeq
with
\beq
 \alpha_{\Psi} = \frac{ 2 Q_{\Psi}^2 M_{\rm P}^2}{m_{\Psi}^2} = 2 \left( \frac{ \partial \ln m_{\Psi}}{\partial \vp} \right)^2
\eeq
where
\beq
 Q_{\Psi} = M_{\rm P}^{-1} \frac{ \partial m_{\Psi}}{\partial \vp} \; .
\eeq
Typically, the cosmon charge $Q$ for two different particles is not proportional to their mass. For example, the proton and neutron cosmon charges $Q_p$ and $Q_n$ will in general not obey $Q_p / m_p = Q_n / m_n$, and similarly for atoms of different composition. As a result, two test bodies of 
different composition with masses $M_b$, $M_c$ will undergo a differential acceleration towards any given gravitational source
\beq
 \eta = 2 \frac{a_b - a_c}{a_b + a_c}\,.
\eeq
We now consider how this experimentally observable quantity can be related to the parameters of each unified scenario, and to cosmological evolution.

\section{Apparent violation of the Weak Equivalence\\ Principle}
\label{sec:WEP}
\subsection{Cosmon charge}
We may define the coupling of a dimensionless scalar $\vp \equiv \Phi / M_{\rm P}$ with kinetic term $M_{\rm P}^2 (\partial \vp)^2/2$ to a source or test body of mass $M_b$ 
as 
\beq 
 \lambda^b \equiv \frac{\partial}{\partial \vp} \ln \frac{M_b}{M_{\rm P}} = \frac{M_{\rm P}{M_b}}{Q_b}\,.
\eeq
Note that $\lambda^b$ is the analogue of $\beta_k$, Eq.~\eqref{eq:DefBetak}, if we take dimensionless mass ratios involving the Planck mass (which is fixed in the Einstein frame). Given a gravitational source of mass $M_s$ separated by a distance $r$ from two test bodies $M_b$, $M_c$, the static configuration for a scalar of mass $m_\vp\ll 1/r$ is given by solving the Poisson equation with a source term proportional to $\lambda^s$ 
supported within the source body. The resulting gradient of $\vp$ at large distance is proportional to 
the gravitational field of the source $g$, thus $\partial \vp/\partial r = 2\lambda^s g$ up to a sign and $\Delta\vp = 2\lambda^s\Delta U$ for small variations.~\footnote{The factor 2 arises from the normalization of $\vp$ via the reduced Planck mass $1/\sqrt{8\pi G}$.} Spatial variations of couplings are given by Eq.~(\ref{DlnGdvphi}) as $\Delta \ln G_i = d_i\beta_X \Delta\vp$, thus the ``coupling coefficients'' $k_i\equiv \Delta \ln G_i/\Delta U$ used in \cite{DentEotvos} are given by $k_i = 2 d_i \beta_X \lambda^s$. As discussed there, the resulting spatial variations within the Solar System are in principle directly measurable via atomic clocks, however considerable improvements in sensitivity would be required for such measurements to compete with current WEP bounds.

A test body will experience accelerations towards the source, due to the dependence of its mass on $\vp$, of magnitude $2g\lambda^b\lambda^s$. Thus the E{\" o}tv{\" o}s parameter $\eta$ measured by WEP experiments is given by \cite{Wetterich:2002ic,Dent06}
\beq \label{etadef}
 \eta^{b-c} \equiv 2\frac{a_b-a_c}{a_b+a_c} \simeq 2\lambda^s(\lambda^b-\lambda^c),
\eeq
where $a_{b,\,c}$ are the accelerations towards the source 
of the two test masses. This may also be derived via ``sensitivity coefficients'' 
\beq
  \lambda_i^{b-c} = \frac{\partial \ln (M_b/M_c)}{\partial \ln G_i} 
\eeq
defined in \cite{DentEotvos}, such that $\eta^{b-c}=\sum_i k_i \lambda^{b-c}_i$. We then expand~\footnote{We use the Einstein summation convention from now on unless stated otherwise.}
\beq \label{lambdablambdac}
 \lambda^b - \lambda^c = \frac{\partial }{\partial \vp} \ln \frac{M_b}{M_c}
 = \lambda_i^{b-c} 
\frac{\partial \ln G_i}{\partial \vp}
 = \lambda_i^{b-c} d_i \beta_X = \frac{\lambda_i^{b-c} k_i}{2\lambda^s}.
\eeq
The relevant couplings $\lambda^b$ or cosmon charges $Q_b$ may be derived from the dependence of the mass of atoms 
on $\vp$. It is convenient to consider the mass per nucleon, thus 
\begin{multline}
 \lambda^b = \frac{\partial}{\partial \vp} \ln  \frac{M_b}{AM_{\rm P}} \nonumber \\ 
 \simeq \frac{\partial}{\partial \vp} \ln \left[M_{\rm P}^{-1} 
 \left(\!m_N - \left(\!R_Z-\frac{1}{2}\!\right) \delta_N + R_Z m_e - \frac{B(A,Z)}{A} \right)\right] \\
 \simeq 
 \partial_\vp \ln \frac{m_N}{M_{\rm P}}
 -\left(\!R_Z-\frac{1}{2}\!\right)\frac{\delta_N}{m_N} 
 \partial_\vp \ln \frac{\delta_N}{m_N} +R_Z\frac{m_e}{m_N} 
 \partial_\vp \ln \frac{m_e}{m_N} -\frac{B}{Am_N} 
 \partial_\vp \ln \frac{B/A}{m_N},
\end{multline}
where $m_N=(m_p+m_n)/2$, $\delta_N=m_n-m_p$, $R_Z\equiv Z/A$ is the proton fraction and $B$ is the nuclear binding energy. 

The value of $\lambda^s$ is typically dominated by the dependence of $m_N/M_{\rm P}$ on $\vp$. For unified scenarios it can be related to the $\vp$-dependence of $l$ using results from \cite{Part1} as
\beq
 \lambda^s = \frac{\partial}{\partial l}\ln\frac{m_N}{M_{\rm P}}\!\cdot\! \frac{\partial l}{\partial \vp} 
 \equiv d_{Ng} \frac{\partial l}{\partial \vp}
 \simeq \left( d_M + 0.58\frac{d_X}{\alpha_X} +0.35 d_H +0.37 d_S \!\right) \frac{\beta_X}{d_X},
\eeq
where, if $\alpha_X$ varies with $\vp$, we take $l =\ln \alpha_X$ and $d_X=1$. If  $\alpha_X$ does not vary we replace $\beta_X/d_X$ by $\beta_H/d_H$ and choose $d_H=1$.~\footnote{If we take $\alpha$ alone to vary, the only contribution to $\lambda^s$ is from the electromagnetic contributions to the proton mass and nuclear binding energies, hence we evaluate this case separately.}

For ($\lambda^b - \lambda^c$) the leading term $\partial \ln ( m_N / M_{\rm P} ) / \partial \vp$ cancels. The differential coupling then results from the isospin-violating couplings to $\delta_N$ and $m_e$, and differences in binding energy per nucleon. This last term $B/A$ contains an electromagnetic contribution proportional to $\alpha Z(Z-1)/A^{4/3}$, and a dominant contribution of strong nuclear forces, which to first approximation does not depend on $A$. However, in \cite{DentEotvos} the effects of subleading strong interaction terms in $B/A$, principally the surface term $-a_S A^{-1/3}$,
were included and found to be significant if the light quark mass $\hat{m}/\Lambda_c$, and thus the pion mass, varies. 

To evaluate $\lambda^b - \lambda^c$ we expand as Eq.~(\ref{lambdablambdac}) using the 
``nuclear parameters'' $G_I=\{E_n/m_N$, $\alpha$, $a_{nuc}/m_N\}$, where $E_n=\delta_N-m_e$, and $a_{nuc}\propto a_V \propto a_S$ is a parameter giving the mass scale of strong nuclear binding energy.
In terms of these parameters 
we find
\beq
 \lambda_I^{b-c} = -(R_Z^b - R_Z^c) \frac{ \partial}{\partial \ln G_I} \left( \frac{E_n}{m_N} \right) - \frac{ \partial}{\partial \ln G_I} \left( \frac{B^b}{A^b m_N} - \frac{B^c}{A^c m_N} \right).
\eeq
With $\Delta R_Z = R_Z^b - R_Z^c$ this yields
\beq \label{lambdaEn}
 \lambda_{E_n}^{b-c} = - \Delta R_Z \frac{E_n}{m_N},
\eeq
while the dependence on the other nuclear parameters is given by the binding energy and is discussed in \cite{DentEotvos}.

In a given unified scenario the dependences of $G_I$ on $\vp$ variation arise via
\beq
  \frac{ \partial \ln G_I}{\partial \vp} = d_I \frac{ \partial l}{\partial \vp},
\eeq
and one finds
\begin{align}
	d_{En}     &\simeq -\frac{1.29}{\alpha_X} d_X + 1.41 d_H -0.82 d_S, \nonumber \\
	d_{\alpha} &= \frac{80\alpha}{27 \alpha_X} d_X +\frac{43}{27}\frac{\alpha}{2\pi} d_H
	+ \frac{257}{27}\frac{\alpha}{2\pi} d_S, \nonumber \\
	d_{a,nuc} &\approx -0.9 \frac{\Delta \ln (\hat{m}/\Lambda_c)}{\Delta \ln \alpha_X}
	\simeq \frac{0.63}{\alpha_X} d_X - 0.7 d_H + 0.4 d_S, 
\end{align}
where Eqs.~(24), (25) and (44) of \cite{Part1} were used for the dependence of $E_n$, $a_{nuc}$, and
$m_N/M_{\rm P}$ on fundamental parameters. 

In any given unified model the couplings $\lambda^s$, $\lambda^{b,c}$ are proportional to the coupling of $\vp$ to the fundamental parameter. For a nonzero $\beta_X$ ($\beta_H$)
we thus have $\eta\propto \beta_X^2$ ($\beta_H^2$) and the fundamental coupling at the present value of $\vp$ is directly bounded.

\subsection{Unification factor}
\label{sec:UnificationFactor}

We would like to compare the differential acceleration $\eta$ with the time variation of couplings. In order to arrive at the simple formula \eqref{eq:DifferentialAccelerationParam}, we first consider a ``fiducial unified scenario'' for which the unification factor satisfies $F=\Delta R_Z$ in this equation. Other unified scenarios can then be characterized by the difference of $F/\Delta R_Z$ 
from unity. 

As fiducial unified scenario we consider a variation of $\alpha_X$ in a non-SUSY GUT ($\alpha_X\simeq 1/40$), accompanied by a variation of $\vev{\phi}$ such that the ratio between the Fermi scale $\vev{\phi}$ and the QCD scale $\Lambda_c$ remains constant. This corresponds to the unification coefficients
\beq \label{fiducialdvalues}
 d_M = 0, \; d_X = 1, \; d_H = \frac{2\pi}{7\alpha_X} \simeq 36, \; d_S = 0 .
\eeq
We also neglect for the fiducial calculation the contributions from nuclear binding energy. These can be incorporated into the unification factor via
\beq \label{FCDDprime}
 F = C_{GUT} \left( \Delta R_Z + D_{GUT} \Delta \frac{Z(Z+1)}{A^{4/3}} + D'_{GUT} \Delta A^{-1/3} \right),
\eeq
where $C_{GUT}$ accounts for the source coupling and value of $d_{En}$, and the terms with prefactors $D_{GUT}$ and $D'_{GUT}$ account for nuclear Coulomb self-energy and nuclear surface energy, respectively.

For the fiducial calculation the preceding relations 
yield
\beq
 \eta^{b-c} = 2 \lambda^s \Delta R_Z \frac{-E_n}{m_N} d_{E_n} 
 \frac{ \partial l}{\partial \vp}
 = 2 \frac{0.90}{\alpha_X} \Delta R_Z \frac{-E_n}{m_N} (-0.97 d_\alpha) 
 \left( \frac{ \partial \ln \alpha_X}{\partial \vp} \right)^2\, ,
\eeq
using the relation $d_{En}\simeq -0.97d_{\alpha}$ that holds if fermion masses do not vary relative to $\Lambda_c$.
Now writing
\beq
 \frac{\partial\ln \alpha_X}{\partial \vp} = \frac{1}{\dot{\vp}} 
 \frac{\partial\ln \alpha_X}{\partial t} 
 = \frac{d_\alpha^{-1}}{\dot{\vp}} \left( \frac{\dot{\alpha}}{\alpha} \right) 
 = \frac{2877\alpha_X}{67} \frac{1}{\dot{\vp}} \left( \frac{\dot{\alpha}}{\alpha} \right),
\eeq
and using the cosmological relation $\dot{\vp}^2 = 3H_0^2 \Omega_h(1+w_h)$ we arrive at
\beq \label{eq:Fiducialeta}
 \eta \simeq 3.8 \times 10^{-12} \left( \frac{\dot{\alpha}/\alpha}{10^{-15} {\rm y}^{-1}} \right)^2 \frac{ \Delta R_Z }{\Omega_h (1 + w_h)}\,. 
\eeq
Here
$\Omega_h$ denotes the fraction of the energy density of the scalar field as compared to the total energy density of the Universe. The equation of state $w_h = (T-V)/(T+V)$ can range between -1 and +1 for a single scalar field. If the scalar is responsible for the dark energy in the Universe, one needs $\Omega_h \approx 0.73$ and $1 + w_h \lesssim 0.1$ \cite{WMAP5}. The relation \eqref{eq:Fiducialeta} is also independent of the value of $\alpha_X$, as long as Eq.~\eqref{fiducialdvalues} is satisfied.

We next consider the contributions from Coulomb self-energy 
and nuclear surface energy. 
The ratio of the Coulomb term to the $\Delta R_Z$ term in $\eta$ yields in general
\beq
 D_{GUT} =  \frac{a_C}{-E_n} \frac{d_\alpha}{d_{En}} \simeq
 -0.91\frac{\frac{80\alpha}{27 \alpha_X} d_X +\frac{43}{27}\frac{\alpha}{2\pi} d_H
	+ \frac{257}{27}\frac{\alpha}{2\pi} d_S}{-1.29\alpha_X^{-1} d_X + 1.41 d_H -0.82 d_S},
\eeq
where $a_C\simeq 0.71\,$MeV. In our fiducial unified scenario the strong binding energy coefficient $a_{nuc}/m_N$ does not vary, thus $D_{GUT}\simeq a_C/(0.97E_n)\simeq 0.94$ and $D'_{GUT}=0$. More generally we have
\beq
 D'_{GUT} = \frac{a_S}{-E_n}\frac{d_{a,nuc}}{d_{En}}  
 \simeq -22.8 \frac{0.63\alpha_X^{-1} d_X - 0.7 d_H + 0.4 d_S}{-1.29 \alpha_X^{-1} d_X + 1.41 d_H -0.82 d_S},
\eeq
where $a_S\simeq 17.8\,$MeV.

By definition the fiducial scenario has $C_{GUT} = 1$. In a general unified scenario we insert the appropriate values for the unification coefficients $d_k$,
\beq
 C_{GUT} = \frac{d_{Ng}}{d_{Ng(0)}} \frac{ d_{E_n}}{d_{E_n(0)}}
 \left(\frac{d_{\alpha(0)}}{d_\alpha}\right)^2,
\eeq
where $\lambda^s_{(0)}$, $d_{E_n(0)}$ and $d_{\alpha(0)}$ are the values for the fiducial scenario. 
In Table~\ref{tab:CgutDgut} we display $C_{GUT}$, $D_{GUT}$ and $D'_{GUT}$ for the various unification scenarios. 
\begin{table}
\center
\begin{tabular}{c||c|c|c|c|c|c|c}
\hline
Scenario & 5, $\tilde{\gamma}\! =\! \frac{2\pi}{7\alpha_X}$ & 2 & 3 & 4 & 5, $\tilde{\gamma}\! =\! 42$ & 6, $\tilde{\gamma}\! =\! 70$ & 6, $\tilde{\gamma}\! =\! 25$ \\ 
\hline 
\hline 
$C_{GUT}$  & 1    & 43    & -3900  & -69    & -8.8  & -9.0  & 19 \\
$D_{GUT}$  & 0.94 & 0.015 & -0.001 & -0.020 & -0.11 & -0.12 & 0.048 \\
$D'_{GUT}$ & 0    & 11.1  & 11.3   & 11.5   & 12.7  & 12.8  & 10.7 \\
\hline
$F$ (Be-Ti)&-1.90 & 95    & -9000  & -165   & -25   & -26   & 41 \\
\hline
\end{tabular}
\caption{Values of $C_{GUT}$, $D^{(')}_{GUT}$ for the ``fiducial scenario'' and for unified scenarios considered in \cite{Part1}; we also give values of $F$ for a WEP experiment using Be-Ti masses.} 
\label{tab:CgutDgut}
\end{table}
We also display the values of $F$ obtained by combining the different ingredients in Eq.~(\ref{FCDDprime}), and considering Be and Ti test masses ($\Delta R_Z \simeq -0.015$, $\Delta Z(Z-1)A^{-4/3}\simeq -2.0$, $\Delta A^{-1/3}\simeq 0.21$).
The very large value of $F$ for Scenario 3 reflects an accidental cancellation of contributions to $\partial \ln \alpha/\partial l$ which we do not consider to be typical.

\subsection{WEP limits on variations at the present epoch}

The time evolution of a scalar contributes to the effective cosmological equation of state via its kinetic term to bound $\dot{\vp}$ at present, in terms of the energy density fraction $\Omega_h$ and equation of state $w_h$ of quintessence. Previously \cite{Dent06} we took a conservative upper limit of $\Omega_h(1+w_h) \le 0.16$ leading to a ``speed limit'' $\dot{\vp} \leq \dot{\vp}_{\rm max}\simeq 5\times 10^{-11}\, \rm{y}^{-1}$ for a Hubble rate of $7\times 10^{-11}\, \rm{y}^{-1}$. 
Considering recent improvements in cosmological measurements we may now take $w_h\leq -0.9$ and $\Omega_h\simeq 0.73$ at the present epoch (if the evolution of the field has not accelerated at late times), thus $\dot{\vp}/H_0\leq 0.47$ and we have
\beq \label{speedlimit}
 \dot{\vp}_{\rm max}\simeq 3.5\times 10^{-11}\, \rm{y}^{-1}.
\eeq
Combining this with the upper limit from WEP on the fundamental coupling $\beta_X$ or $\beta_H$ in each unified model, we bound the present rate of variation of the fundamental parameter $\alpha_X$ or $\vev{\phi}/M_X$. The bounds can be extrapolated to past epochs by expanding the coupling functions and the scalar time evolution about the present values, if these are smooth and not too rapidly-varying functions. 

The experiment setting the tightest limits on scalar couplings \cite{Schlamminger:2007ht} has the result
\beq \label{Schlameta}
 \eta = (0.3\pm 1.8)\times 10^{-13}
\eeq
for test bodies of Be 
and Ti 
composition, where the gravitational source is taken to be the Earth. 

\paragraph{Variation of $\alpha$ alone}
Here we parameterize the variation as $\Delta \ln \alpha = \beta_\alpha \Delta \vp$, and we have for the source coupling 
\beq
\lambda^s = \left( 3\times 10^{-4}+ 8\times 10^{-4}(R_Z-1/2) + \frac{a_C}{m_N} \frac{Z(Z-1)}{A^{4/3}} \right) \beta_\alpha
\eeq 
where the coefficients arise from the contributions of electromagnetic self-energy to $m_N$, $\delta_N$ and $B/A$ respectively. We have $d_{En}\simeq -0.97$, and clearly also $d_\alpha \equiv 1$, $d_{a,nuc}=0$. The bound arising from Eq.~(\ref{Schlameta}), treating the Earth as composed of iron or silicon (with similar results), is $\beta_\alpha\leq 1.8\times 10^{-4}$ at $1\sigma$. Thus the present rate of variation is limited to $|\dot{\alpha}/\alpha| \leq 6.2\times 10^{-15}\, {\rm y}^{-1}$, a somewhat looser bound than from atomic clocks. 

\paragraph{WEP bounds for unified models}
If the unified coupling $\alpha_X$ varies then our fundamental coupling is $\beta_X$. For Scenario 2 we find the bound 
\[
 \beta_X^2 \leq 6.2\times 10^{-11}\alpha_X^2\qquad (1\sigma)
\]
thus the present rate of variation is limited to 
\beq
 |\dot{\alpha}_X/\alpha_X| \leq 2.8\times 10^{-16}\alpha_X\,{\rm y}^{-1},
\eeq
equalling $1.2\times 10^{-17}\,{\rm y}^{-1}$ for SUSY theories ($\alpha_X\simeq 1/24$) or $6.9\times 10^{-18}\,{\rm y}^{-1}$ for non-SUSY ($\alpha_X\simeq 1/40$). A major contribution to $\eta$ is the strong nuclear binding energy with coefficient $a_S/m_N$, which is sensitive to variation of the light quark mass relative to the QCD scale. 

For Scenarios 3 and 4 we take instead a fundamental coupling $\beta_H$ describing the variation of $\vev{\phi}/M_X$ (and $\tilde{m}/M_X$ in the SUSY case).
In Scenario 3 the magnitude of
$\beta_H$ is limited to $9.5\times 10^{-6}$ ($1\sigma$) and time variation is bounded by $|\partial_t \ln (\vev{\phi}/M_X)|\leq 3.3\times 10^{-16}$.
For Scenario 4 the bounds are similar with $\beta_H,\,\beta_S\leq 1.0\times 10^{-5}$ and $|\partial_t \ln (\vev{\phi}/M_X)|\leq 3.5\times 10^{-16}$.

For Scenarios 5 and 6 we take again a unified coupling $\beta_X$:
in Scenario 5 with $\tilde{\gamma}=42$ we find an upper limit $|\beta_X|\leq 3.6\times 10^{-7}$, thus $\dot{\alpha}_X/\alpha_X$ is bounded at $1.3\times 10^{-17}\,{\rm y}^{-1}$ at $1\sigma$.

In Scenario 6 with $\tilde{\gamma}=70$ the relevant $1\sigma$ bounds are still tighter with $|\beta_X|\leq 2.3\times 10^{-7}$ and $|\dot{\alpha}_X/\alpha_X| \leq 8.1 \times 10^{-18}\,{\rm y}^{-1}$. With $\tilde{\gamma}=25$ here we obtain $|\beta_X|\leq 3.1\times 10^{-7}$ and $|\dot{\alpha}_X/\alpha_X| \leq 1.1 \times 10^{-17}\,{\rm y}^{-1}$.

\paragraph{Cancellations in WEP bounds}
For particular values of $\tilde{\gamma}$ in Scenarios 5 and 6 there may be cancellation between different contributions to $\eta$, thus $\lambda_b$ and $\lambda_c$ may happen to be nearly equal for the test bodies considered. Without SUSY (Scenario 5) this happens for $\tilde{\gamma}\simeq 35$, approximately the value $80\pi/7$ at which the variation of $\vev{\phi}/\Lambda_c$ also vanishes.
With SUSY (Scenario 6) and $d_S$ set equal to $d_H$, the analogous value of $\tilde{\gamma}$ for which $\Delta \ln \vev{\phi}/\Lambda_c=0$ is $16\pi\simeq 50$. 
Near these values, contributions 
from variation of fermion masses relative to $\Lambda_c$ become small, leaving the intrinsically smaller effect of varying $\alpha$. In principle one might fine-tune $\tilde{\gamma}$ to produce a yet more dramatic cancellation in any given experiment. However the cancellation cannot be exact for all pairs of test bodies, thus bounds on $\beta_X$ from all WEP experiments cannot be removed by a fine-tuned choice of $\tilde{\gamma}$, only slightly weakened.

\subsection{Bounds on present-day variation and testing unified\\ scenarios} \label{present}

Within our theoretical framework there exist three distinct ways to bound or measure the present-day rate of variation of fundamental parameters. The first is a direct measurement, for instance atomic clock experiments. If one or more nonzero variations is found in this way, bounds on unified models may immediately be set. The second method is by combining information on the size of scalar field couplings from WEP tests (Section~\ref{wep}) with a cosmological upper bound on the kinetic energy of scalar fields \cite{Wetterich:2002ic,Dent06}. Such bounds on scalar couplings will depend on the choice of unified model and in general will be independent of those derived from atomic clocks. 

Thirdly, under the assumption of a monotonic variation (that also does not significantly accelerate with time), we may convert any ``historic'' bound on the net variation of a fundamental parameter since a given time, into a bound on the present rate of variation:
\beq \label{historic}
	|\dot{G}_k| \leq (t_0-t_n)^{-1} \left| G_k(t_0)-G_k(t_n)\right| \equiv \frac{|\Delta G_k|}{\Delta t},\qquad t_n < t_0,
\eeq
where $t_0$ denotes the present time. 

For any given unified model of time variations, the three methods to bound present-day evolution will have different sensitivity. Therefore if one method gives a clear nonzero variation we would (in some cases) be able to distinguish between models, given that the other bounds are still consistent with zero. To give a simple example, the direct detection of a nonzero time variation in atomic clocks near the present upper bound would immediately rule out a large class of models that cannot account for such a variation without leading to WEP violation above current limits, and would also rule out models in which such nonzero variations extrapolated to past epochs $t_n<t_0$ would exceed observational bounds.

However, such inferences do not function equally in all directions. A nonzero finding of WEP-violating differential acceleration would indicate nontrivial scalar couplings, but need not imply nonzero time variation since the rate of change of the scalar is not bounded below. Also, a nonzero variation at some past epoch $t_n$ would not necessarily imply a lower bound to the present-day rate of variation or size of scalar couplings, since the variation could have slowed substantially since then (either due to nonlinear scalar evolution or a nonlinear coupling function). With the assumption of a smooth and monotonic variation of the scalar field and its coupling functions, one could estimate, for any given unified model, where the first signals of present-day or recent variation are expected to appear. 

At present these methods give null results up to redshifts about $0.8$, but if a nonzero time variation exists, we can determine for each unified scenario which observational method is most sensitive. Thus if a nonzero signal of recent variation arises it may be used to distinguish between models. 
We assume for this purpose an approximately linear variation over recent cosmological times, thus measurements of absolute variation at nonzero redshift $z$ imply time derivatives
\beq
  \frac{d \ln X}{dt} \simeq \frac{\Delta \ln X (z)}{t_0 - t(z)}.
\eeq
Here $X$ is the fundamental varying parameter: we consider first $X\equiv \alpha$, if the only varying parameter is $\alpha$; in scenario 1, $X\equiv G_{\rm N}m_N^2$, in scenarios 2, 5 and 6, $X \equiv \alpha_X$ and in scenarios 3 and 4, $X \equiv \vev{\phi}/M_X$. Then Table~\ref{tab:PresentVariationErrors} gives the precision of bounds on time derivatives for the unified scenarios we consider.~\footnote{Except scenario 1 (varying $G_N$) which is probed by a quite different set of measurements.} As in \cite{Part1} we take the Oklo bound as applying directly to the variation of $\alpha$, and increase its uncertainty by a factor 3 to account for possible cancellations when other parameters also vary. We present the Rosenband {\it et al.}\ \cite{Rosenband} Al/Hg ion clock bound separately to illustrate to what extent it improves over previous atomic clock results.
\begin{table}
\center
\begin{tabular}{|lc|cccccc|}
\hline
Scenario & $X$              & Clocks         & Al/Hg  & WEP & Oklo $\alpha$ & Meteorite & Astro
\\
\hline
$\alpha$ only & $\alpha$    & 0.13 ($\alpha$)& 0.023 & 6.2   & 0.033 & 0.32   & 0.44 ($y$) \\
2        & $\alpha_X$       & 0.074 ($\mu$)  & 0.027 & 0.007 & 0.12  & 0.015 & 0.006 (NH$_3$) \\
2S       & $\alpha_X$       & 0.12  ($\mu$)  & 0.044 & 0.012 & 0.19  & 0.026 & 0.010 (NH$_3$) \\
3        & $\vev{\phi}/M_X$ & 2.6 ($\mu$)    & 12.4  & 0.33  & 54    & 0.53  & 0.22  (NH$_3$) \\
4        & $\vev{\phi}/M_X$ & 6.2 ($\mu$)    & 1.78  & 0.35  & 7.7   & 1.2   & 0.51  (NH$_3$)
\\
5, $\tilde{\gamma} = 42$ & $\alpha_X$ & 0.32 ($\alpha$) & 0.024  & 0.013 & 0.11  & 0.069 & 0.035
(NH$_3$) \\
6, $\tilde{\gamma} = 70$ & $\alpha_X$ & 0.21 ($\alpha$) & 0.016  & 0.008 & 0.070 & 0.049 & 0.025
(NH$_3$) \\
6, $\tilde{\gamma} = 25$ & $\alpha_X$ & 0.25 ($\mu$)    & 0.027  & 0.011 & 0.12  & 0.056 & 0.021
(NH$_3$) \\
\hline
\end{tabular}
\caption{Competing bounds on recent ($z\leq0.8$) time variations in unified scenarios. For each scenario we give $1\sigma$ uncertainties of bounds on $d(\ln X)/dt$ in units $10^{-15}{\rm y}^{-1}$, where $X$ is the appropriate fundamental parameter. The Oklo bound is rescaled in Scenarios 2-6 as explained in the main text. The column ``Clocks'' indicates whether $\alpha$ or $\mu$ gives the stronger bound; the recent Al/Hg limit \cite{Rosenband} is given a separate column. The column ``Astro'' indicates which measurements of astrophysical spectra are currently most sensitive in each scenario. 
}
\label{tab:PresentVariationErrors}
\end{table}

Extending the methods of this section beyond $z \approx 0.5$ becomes questionable. One could use linearity in $\ln(1+z)$ instead of $t$, but even this improvement may lead to unreliable extrapolations for models with a particular dynamics of the scalar field, such as crossover quintessence. An alternative approach to relating present-day variation to cosmological history in previous epochs, under certain assumptions on the scalar evolution, is given in \cite{Avelino:2008dc}.

\section{Monotonic cosmon evolution and crossover\\ quintessence}
\label{sec:QuintessenceModels}

\subsection{Crossover quintessence}
We now consider two illustrative models which predict the evolution of the quintessence field $\vp(t)$, consistently with other cosmological bounds, given a small number of adjustable parameters. Our aim is to see to what extent such models can be consistent with the behaviour of time variations that we have outlined.

Our first class of models is ``crossover quintessence'' \cite{Hebecker:2000zb,Doran:2007ep,Wetterich:2002wm}: here the scalar field follows tracking solutions \cite{Wetterich88, Ratra88} at large redshift. In this early epoch the equation of state $w_h\equiv (T-V)/(T+V)$, where the kinetic energy $T=M_{\rm P}^2 \dot{\vp}^2/2$ and potential energy is $V(\vp)$, is equal to that of the dominant energy component (matter or radiation). At some intermediate redshift before the onset of acceleration, the time evolution of the cosmon slows down. In consequence, there is a crossover to a negative equation of state and the fraction of energy density due to the scalar begins to grow. In recent epochs the field has an effective equation of state $w_h \gtrsim -1$. We do not aim to build and solve detailed models of this type, but rather estimate general properties of the scalar evolution.

We begin with a caricature of a scaling quintessence or early dark energy model, in which the dark energy equation of state is constant at late times at $w_{h0}$. Above some given redshift $z_+$ the equation of state crosses over to the scaling condition $w_h=0$ in the matter dominated era; then for $z>z_{eq}$, before matter-radiation equality, we again have scaling through $w_h=1/3$. Then the general relation ($a = (1+z)^{-1}$)
\beq
	\frac{d \ln\rho}{d \ln a} = -3(1+w(a))
\eeq
may be used to find the matter, radiation and dark energy densities over cosmological time. We estimate the scalar kinetic energy via $M_{\rm P}^2 \dot{\vp}^2/2=\rho_h(1+w_h)/2$ and can thus integrate $d\vp/da=\dot{\vp}/aH$ from the present back to any previous redshift. The initial conditions are set by specifying the present densities of matter, radiation and dark energy and the model parameters $w_{h0}$ and $z_+$.

Assuming a constant coupling $\delta$ to the fundamental varying parameter, usually $\alpha_X$, the variation is given by
\beq \label{eq:DalphaXgrowing}
	\Delta \ln \alpha_X(z) = \delta( \vp(z) - \vp(0)).
\eeq
This ansatz implies a monotonic evolution, since the increase of $\vp$ with time is monotonic for crossover quintessence. We discussed the viability of monotonic evolution in \cite{Part1} where a first judgement can be made by inspection of figs~1-6 for the various unification scenarios, or Table~\ref{tab:multiplierFactors} in the appendix of the present paper.

To allow us to compare easily with the observational results, we observe that in Table~\ref{tab:multiplierFactors} the constraints for $l_5$ are considerably weaker than those for the remaining evolution factors. Furthermore, the one-sigma range for $l_4$ is nonzero for all scenarios.
Hence we shall compare observational and theoretical values for the ratios $l_1/l_4$, $l_2/l_4$, $l_3/l_4$ and $l_6/l_4$. We note the opposite sign of $l_3$ and $l_4$ for scenarios 2 and 3, and scenario 6 with $\tilde{\gamma}=25$, which disfavours any monotonic evolution for these scenarios. The averaged observational values in each epoch are given in Table~\ref{tab:RatioMultiplierFactors}.
\begin{table}
\center
\begin{tabular}{|c|cccc|}
\hline
Scenario & $l_1/l_4$ & $l_2/l_4$ & $l_3/l_4$ & $l_6/l_4$\\
\hline
0   & $  0.00 \pm 0.01$ & $0.13 \pm 0.13$ & $0.31  \pm 0.18$ & $-600 \pm 4000$\\
2   & $ -0.10 \pm 0.12$ & $0.04 \pm 0.04$ & $-1.59 \pm 0.88$ & $-380 \pm 230$ \\
3   & $ -0.11 \pm 0.13$ & $0.04 \pm 0.03$ & $-0.12 \pm 0.91$ & $-390 \pm 240$ \\
4   & $ -0.04 \pm 0.10$ & $0.04 \pm 0.03$ & $0.97  \pm 0.64$ & $-380 \pm 260$ \\
5, $\tilde{\gamma} = 42$  & $ 0.00 \pm 0.03$ & $0.04 \pm 0.03$ & $0.41 \pm 0.19$ &
$-280 \pm 191$ \\
without BBN               &          ''       &       ''        &        ''        &
$-66 \pm 165$  \\
6, $\tilde{\gamma} = 70$  & $ 0.00  \pm 0.03$ & $0.02 \pm 0.02$ & $0.39  \pm 0.17$ &
$-275 \pm 150$ \\
without BBN               &          ''       &       ''        &        ''        &
$-69 \pm 139$  \\
6, $\tilde{\gamma} = 25$  & $ -0.04 \pm 0.06$ & $0.02 \pm 0.04$ & $-1.04 \pm 0.49$ &
$-343 \pm 142$ \\
with Wendt               & $ 0.03  \pm 0.05$ & $-0.01\pm 0.03$ & $0.70 \pm 0.52$  & 
$231 \pm 163$  \\
\hline
\end{tabular}
\caption{Ratios of the evolution factors from observations.}
\label{tab:RatioMultiplierFactors}
\end{table}

Note that a constant coupling $\delta$ drops out of the ratios $l_i/l_j$. Thus we may probe cosmon/quintessence evolution directly without knowing the absolute size of the coupling, only assuming its (approximate) constancy over the relevant range of evolution. Due to its monotonic evolution, crossover quintessence cannot give negative ratios $l_i/l_j$. Hence it cannot be a good fit to the \lise\ abundance within the unification scenarios 2 to 6 that we consider. This reflects the tension between the \lise\ abundance and a positive variation of $\mu$ discussed in Section~4.4 of \cite{Part1}. 

In Table~\ref{tab:TheoreticalRatios} we display the values of the ratios $l_i/l_4$ and $l_i/l_3$ expected from crossover quintessence for various values of the parameters $w_{h0}$ and $z_{+}$. Considering the \lise\ abundance to be affected by astrophysical systematics in scenarios 5 and 6 ($\tilde{\gamma} = 70$), or using the null result of \cite{Wendt08} for $\Delta \ln \mu$ at intermediate redshift for scenario 6 ($\tilde{\gamma} = 25$ ``with Wendt''), we conclude that crossover quintessence could, in pinciple, reconcile a coupling variation of the claimed size in epochs 3 and 4 with the bounds from late cosmology, {\em i.e.}\ epochs 1 and 2. This is due to the ``slowing down'' of the cosmon evolution, as noted in \cite{WetterichCrossoverQ}. Values of the present equation of state $w_{h0}$ quite close to $-1$ would be required, however. An observation of coupling variations would put strong bounds on the dynamics of the cosmon field and provide an independent source of information about the properties of dark energy.
\begin{table}
\center
\begin{tabular}{|cc|cccc|}
\hline
$w_{h0}$ & $z_{+}$ & $l_1/l_4$ & $l_2/l_4$ & $l_3/l_4$ & $l_6/l_4$ \\
\hline
-0.95    & 3 & 0.12 & 0.38 & 0.67 & 26\\
-0.99    & 3 & 0.09 & 0.28 & 0.55 & 38\\
-0.9999  & 3 & 0.02 & 0.10 & 0.37 & 54\\
-0.95    & 7 & 0.15 & 0.47 & 0.80 & 13\\
-0.99    & 7 & 0.15 & 0.47 & 0.80 & 24\\
-0.9999  & 7 & 0.08 & 0.27 & 0.52 & 120\\
-0.999999& 7 & 0.02 & 0.09 & 0.34 & 180\\
\hline
\end{tabular}
\caption{Ratios of evolution factors from crossover quintessence. The $l_i$ are evaluated by averaging over the variations evaluated at the same redshift as the data in each epochs (weighting by the number of absorption systems if appropriate).}
\label{tab:TheoreticalRatios}
\end{table}

Note that observational probes of dark energy would {\em not}\/ give results for $w_{h0}$ that coincide with the values that we take in our model. Such probes do not actually measure the present-day equation of state, rather they estimate $w_0$ by extrapolating from past epochs under some parameterization.

\section{Models with growing neutrinos and oscillating \\variation}
\label{sec:Growing}

In this section we investigate models that do not obey the proportionality of all coupling variations for all redshifts and do not show a monotonic evolution of the cosmon field. A systematic analysis of all such models seems difficult, and we concentrate on a specific example.

Growing neutrino models \cite{Amendola_grownu, Wett_grownu} explain the value of today's dark energy density by the ``principle of cosmological selection''. The present fraction of dark energy, $\Omega_h^0$, is set by a dynamical mechanism. As soon as the neutrinos become non-relativistic, their coupling to the cosmon triggers an effective stop (or substantial slowing) of the evolution of the cosmon. Before this event, the quintessence field rolls down an exponential potential and thus follows the tracking behaviour described in the preceding section. This mechanism requires a neutrino mass that depends on the cosmon field $\vp$ and grows in the course of the cosmological evolution. The present dark energy density, $\rho_{h0}$, can be expressed in terms of the average present neutrino mass, $m_{\nu}(t_0)$, and a dimensionless parameter $\gamma$ of order unity \cite{Amendola_grownu},
\beq
(\rho_{h0})^{1/4} = 1.07 \left( \frac{\gamma m_{\nu}(t_0)}{\rm{eV}} \right)^{1/4} 10^{-3} \rm{eV}.
\eeq

We follow again our simple proportionality assumption, namely that the cosmon evolution produces a variation $\Delta \ln \alpha_X(z) = \delta (\varphi(z)-\varphi(0))$, with a proportional variation for other couplings according to the unification scenario. This is the only contribution to the variation of the unified coupling $\alpha_X$ and $M_X / M_{\rm P}$. However, a new ingredient is an additional variation of the Higgs v.e.v.~$\vev{\phi}$ with respect to the Planck mass, which only becomes relevant at late time \cite{Wett_grownu}. It is due to the effect of a changing weak triplet operator on the v.e.v.~of the Higgs doublet. If the dominant contribution to the neutrino mass arises from the ``cascade mechanism'' (or ``induced triplet mechanism'') via the expectation value of this triplet, this changing triplet value is directly related to the growing neutrino mass \cite{Wett_grownu}.

\subsection{Stopping and scaling growing neutrinos}
\begin{figure}
\hspace*{-1.5cm}
 \includegraphics[width=7.5cm]{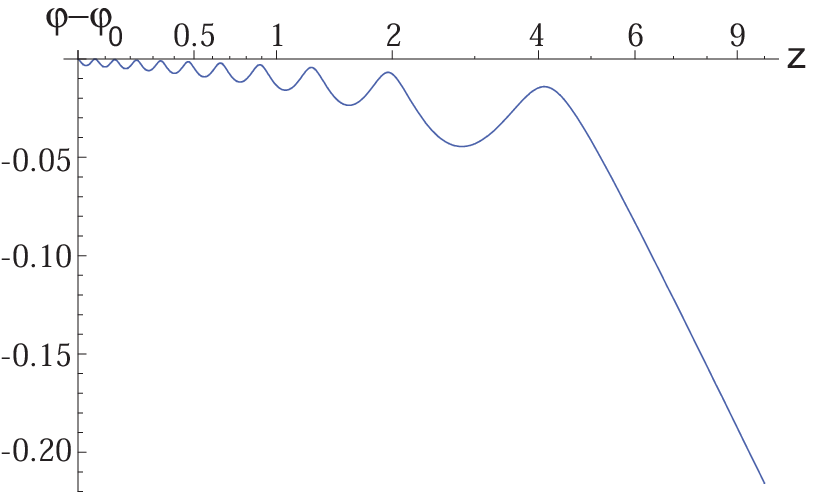} 
\hspace*{-0.2cm}
 \includegraphics[width=8.3cm]{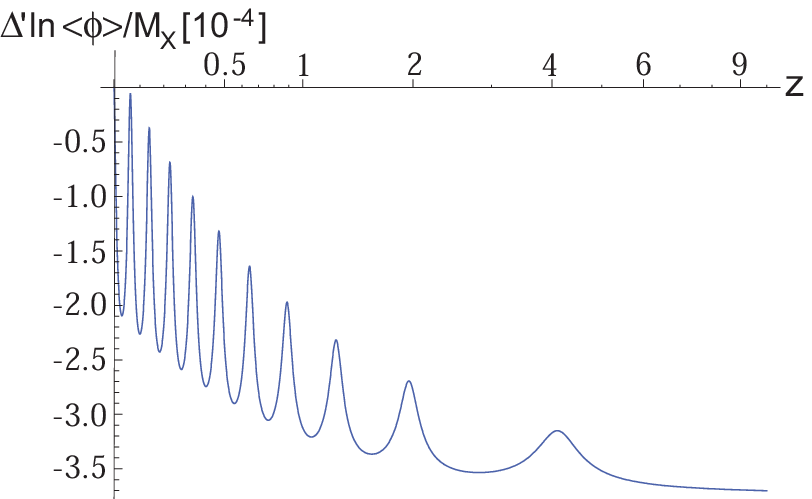}
\caption{Variations in the stopping growing neutrino model of \cite{Wett_grownu}. Left: evolution of the cosmon field. Right: additional variation of the Higgs v.e.v.\ according to Eq.~\eqref{eq:phiRofz}.} \label{fig:VariationFixedPoint}
\end{figure}

We consider two models, with slightly different functional dependence of the Higgs v.e.v.\ and neutrino mass on the scalar field. In the first, where the cosmon asymptotically approaches a constant value (``stopping growing neutrino model'') \cite{Wett_grownu}, the additional Higgs variation is given according to
\beq \label{eq:phiRofz}
	\frac{\vev{\phi}}{M_X}(z) = \bar{H} \left(1 - R(z)\right)^{-0.5},
\eeq
where
\beq \label{eq:RofzphiWett}
	R(z) = \frac{R_0}{1-\exp(-\epsilon(\vp(z) - \vp_t))}.
\eeq
Here, $\vp_t \approx 27.6$ is the asymptotic value (choosing the parameter $\alpha=10$ in the exponential potential \cite{Wett_grownu}). For illustration we take the set of parameters given in \cite{Wett_grownu}, $\epsilon = -0.05$, while $\bar{H}$ is set by demanding the Higgs v.e.v.\ being consistent with measurements today, $\vev{\phi}(z=0) = 175\,$GeV. We set $R_0 = 10^{-7}$, however in general we only require $R(z=0) \ll 1$. The resulting variations are shown in Fig.~\ref{fig:VariationFixedPoint}.

The combined variation is different for each of the unified scenarios. For all unified parameters except $\vev{\phi}/M_X$ we still have a proportionality for the variations at all $z$, since the variations are proportional to $\vp$. The variations due to the direct coupling \eqref{eq:DalphaXgrowing} can be described by our method of evolution factors $l_n$, even though the $l_n$ need not be strictly monotonic due to the oscillations in $\vp$. However, for $\vev{\phi}/M_X$ we now have one variation linearly proportional to the variation of $\vp$, Eq.~\eqref{eq:DalphaXgrowing} and an additional one with a nonlinear dependence, Eq.~(\ref{eq:phiRofz}-\ref{eq:RofzphiWett}). A simple treatment with common evolution factors for all variations will no longer be applicable. Due to the additional $\vp$-dependence of $\vev{\phi}/M_X$ we may have separate evolution factors for $\vev{\phi}/M_X$, different from the (common) evolution factors for the other couplings. 

For example, the ``linear contribution'' \eqref{eq:DalphaXgrowing} may dominate at BBN and induce a positive $l_6$ common for all couplings. In the range $z<10$ the ``non-linear contribution'' \eqref{eq:phiRofz} could be more important, leading to effectively negative $l_{3,4}$ for $\vev{\phi}/M_X$. (Such an effect could, in principle, relieve the tension between \lise\ and a positive $\mu$-variation at high $z$.) In practice, we calculate $\Delta \ln \alpha_X$ and $\Delta \ln \frac{\vev{\phi}}{M_P}$ at each epoch directly from the model, and extract the varying couplings and observables as explained in Section~3 of \cite{Part1}.
Then we may search for a set of parameters $\delta$, $R_0$ which minimizes the $\chi^2$ for all measured variations. 

The stopping growing neutrino model has an oscillation in $\vev{\phi}$ that grows both in frequency and amplitude at late times as $\vp$ approaches its asymptotic value. Such oscillations must not be too strong as measurements between $z=2$ and today would measure a high rate of change. The oscillation may be made arbitrarily small by choosing small $R_0$. The restrictions from the low-z epochs are actually so strong that to a good approximation the non-linear contribution $\sim R_0$ can be neglected. However, the linear variation \eqref{eq:DalphaXgrowing} is independent of $R_0$. It can be described by our method of evolution factors and yields for our set of parameters the ratios 
\bea
  l_1/l_4 &=& 0.008, \nonumber\\
  l_2/l_4 &=& 0.09, \nonumber \\
  l_3/l_4 &=& 0.44, \nonumber \\
  l_6/l_4 &=& 175.
\eea
Comparing this with the numbers given in Table~\ref{tab:RatioMultiplierFactors} shows that this model naturally yields evolution factors which are of the correct order of magnitude. We emphasize that no new parameter has been introduced for this purpose.

The second growing neutrino model \cite{Amendola_grownu} does not lead to an asymptotically constant $\vp$. Now the coupling of the neutrino to the cosmon $\vp$ is given by a constant $\beta$, according to
\beq
	m_\nu = \tilde{m}_\nu e^{-\beta \vp}.
\eeq
This ``scaling growing neutrino model'' leads in the future to a scaling solution with a constant ratio between the neutrino and cosmon contributions to the energy density. 

\begin{figure}
\hspace*{-2cm}
 \includegraphics[width=8cm]{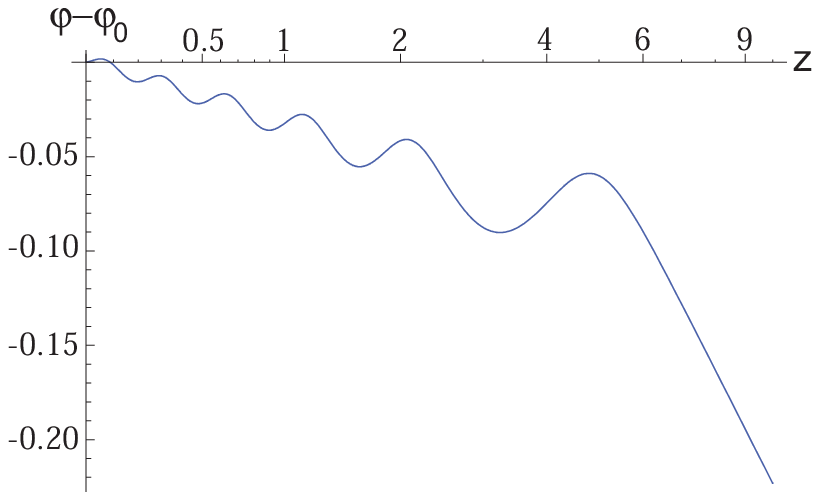} 
\hspace*{0.01cm}
 \includegraphics[width=8.1cm]{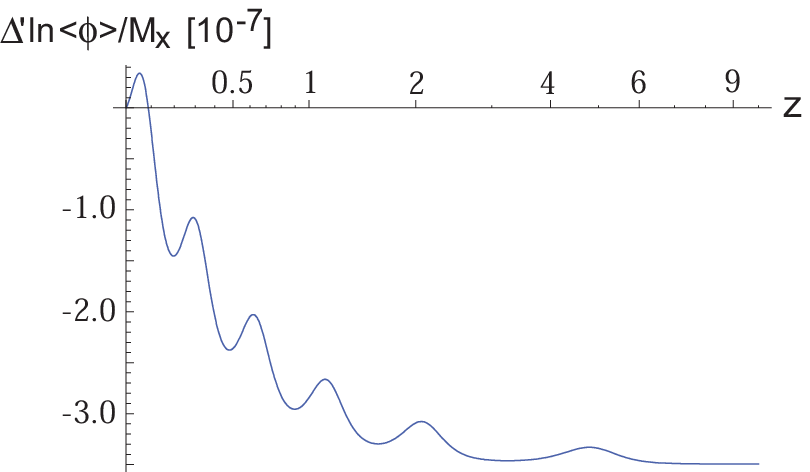}
\caption{Variations in the scaling growing neutrino model of \cite{Amendola_grownu}. Left: evolution of the cosmon field. Right: additional variation of the Higgs v.e.v.\ according to Eq.~\eqref{eq:phiRofz}.} \label{fig:VariationConstantBeta}
\end{figure}

With the choice of parameters $\beta=-52$, $\alpha=10$ and $m_{\nu,0}=2.3\,$eV, and given the triplet mechanism of \cite{Wett_grownu}, the Higgs v.e.v.\ varies as Eq.~(\ref{eq:phiRofz}), where now $R$ is given by 
\beq \label{RofzphiAmen}
	R(z) = R_0 e^{-\beta \vp(z)}.
\eeq
Here the Higgs oscillations remain comparatively small in amplitude, while the absolute value of $\vev{\phi}$ grows overall with time: see Fig.~\ref{fig:VariationConstantBeta} with the parameter choice $R_0=10^{-6}$.
Most of the additional variation of the Higgs v.e.v.\ occurs at later epochs, $z<2$, thus recent observational bounds rule out any significant additional growth in $\vev{\phi}$. We considered fitting the observational values excluding BBN, as a function of the model parameters $\delta$ and $R_0$, and we find always that the value of $R_0$ at the minimum of $\chi^2$ is unobservably small.

\subsection{Global fits to growing matter models}
Each growing neutrino model contains a few parameters that determine the cosmological evolution of the cosmon $\vp$, and the Higgs v.e.v.\ $\vev{\phi}$. Two coupling parameters give respectively the relative strength of variation of $\alpha_X$ with $\vp$, and the relative strength of the additional variation of the Higgs v.e.v.\ due to the varying triplet. For each example of cosmological evolution we may calculate the observables directly in terms of the two coupling parameters and make a global fit for their values. In performing the fit we take the 125 systems of the Murphy {\it et al.} $\alpha$ determination \cite{Murphy:2003mi} within Epochs 3 and 4 (Eq.~(9) of \cite{Part1}) 
and further split them into 5 subsamples each with 25 absorption systems, since the data set extends over a wide range of redshift where there may be significant oscillations.

For the global fits, we take the scaling growing neutrino model \cite{Amendola_grownu} with $\beta = -52$, $\alpha = 10$, $m_{\nu,0} = 2.3\,$eV. With zero variation at all times (no degrees of freedom), we find $\chi^2 = 3.2$ including \lise\ at BBN, and $\chi^2 = 2.4$ neglecting \lise. The results of the best fits with varying couplings are given in Table~\ref{tab:chi2origAmendola}.
\begin{table}[h]
\centering
\begin{tabular}{|l|cc|cc|}
\hline
Scenario        & $\delta \times 10^4$ & $R_0$ & $\chi^2$ & $\Delta \chi^2$\\
\hline
2               &-0.019 & 0.045& 3.1 & 0.2 \\
2 (no \lise)    &-0.040 & 0    & 2.1 & 0.3 \\
3               & 1.6   & 0    & 2.8 & 0.4 \\
3 (no \lise)    & 1.6   & 0    & 2.0 & 0.4 \\ 
4               & 3.8   & 0    & 2.8 & 0.4 \\
4 (no \lise)    & 3.8   & 0    & 2.0 & 0.4 \\
5.42            & 0.30  & 0    & 2.0 & 1.3  \\
5.42 (no \lise) & 0.24  & 0    & 1.9 & 0.5 \\
6.70            & 0.20  & 0    & 1.9 & 1.3  \\
6.70 (no \lise) & 0.16  & 0    & 1.9 & 0.5 \\
6.25            & 0.18  & 0.090& 2.7 & 0.5 \\
6.25 (no \lise) &-0.055 & 0.061& 2.2 & 0.2 \\
\hline
\end{tabular}
\caption{Fitting parameters and minimal $\chi^2$ values for the different unification scenarios
for best fit to the scaling growing neutrino model \cite{Amendola_grownu}. The last column gives the increase in $\chi^2$ produced when $\delta$ and $R_0$ are forced to vanish, {\it i.e.}\ for zero variation.}\label{tab:chi2origAmendola}
\end{table}
No convincing evidence is found for coupling variations within this model. We have investigated some other choices of parameters and also the stopping growing neutrino model, without a substantial change in the overall situation. In view of the unsettled status of the observational data it seems premature to make a systematic scan in parameter space. Our invesigation demonstrates, however, how a clear positive signal for a coupling variation could restrict the parameter space for quintessence models.

\section{Conclusions}
\label{sec:Conclude}
This paper demonstrates how a clear observation of time variation of fundamental couplings would not only rule out a constant dark energy, but also put important constraints on the time evolution of a dynamical dark energy or quintessence.

In Sec.~\ref{sec:WEP} we have seen how the comparison of a varying coupling with the bounds from tests of the equivalence principle can put a lower nonzero bound on the combination $\Omega_h (1+w_h)$, according to
\beq
\label{eq:BoundOnOmegaw}
 \Omega_h (1+w_h) \gtrsim 3.8 \times 10^{18} F (\partial_t \ln \alpha)^2 
 \eta_{\rm max}^{-1} \, ,
\eeq
with $\partial_t \ln \alpha$ in units of y$^{-1}$, and where $\eta_{\rm max} \simeq 1.8 \times 10^{-13}$ is the current experimental limit on the differential acceleration of two test bodies of different composition. Thus, if $|\partial_t \ln \alpha |$ is nonzero and not too small, $w_h$ {\em cannot}\/ be arbitrarily close to $-1$ (a cosmological constant); nor can the contribution of the scalar to the dark energy density be insignificant. 

The precise bound depends on the ``unification factor'' F, which differs between different scenarios of unification and also depends on the composition of the experimental test bodies. We find $1\leq F \leq \rm{few}\times 10^{2}$ for the Be-Ti masses used for the best current bound on $\eta$ and for representative cases of unified scenario.

Conversely, assuming that the scalar field responsible for the varying couplings is the cosmon, whose potential and kinetic energy account for a dynamical dark energy in the Universe, the bounds on $\eta$ can be used to set bounds on the time variation of couplings in the present epoch. For the different unified scenarios we compare the relative sensitivity of these bounds as compared to laboratory measurements or bounds from the Oklo natural reactor and the composition of meteorites in Table~\ref{tab:PresentVariationErrors}.

For a given unified scenario, the bounds on the time variation of various couplings in different cosmological epochs strongly restrict the possible time evolution of the cosmon field, once at least one irrefutable observation of some coupling variation at some redshift becomes available. We have demonstrated this by an analysis that implicitly assumes a nonzero variation, considering both general features and specific quintessence models. We are aware that the actual values for the evolution factors $l(z)$ from this analysis may be premature, since the observational situation is unclear and on moving grounds. For example, taking the recent reanalysis of the variation of the proton to electron mass ratio $\mu$ in Ref.~\cite{King:2008ud} instead of the results in Ref.~\cite{Reinhold:2006zn} used in this paper, would strongly influence the values of the evolution factors. We have demonstrated this in a somewhat different way by investigating the change in the evolution factors if some claimed observations of varying couplings are omitted. Needless to say that without a clear signal any analysis may become obsolete. For the time being our analysis remains a useful tool for the comparison of the relative sensitivity of different experiments and observations, and for a judgement about mutual consistency of different claimed variations and bounds from other observations.

\subsection*{Acknowledgements}
We acknowledge useful discussions with M.~Doran, G.~Robbers, M.~Pospelov, J.~Donoghue and P.~Avelino, and correspondence with V.~Flambaum. T.\,D. thanks the Perimeter Institute for hospitality during the workshop ``Search for Variations in Fundamental Couplings and Mass Scales''. T.\,D. is supported by the {\em Impuls- and Vernetzungsfond der Helmholtz-Gesellschaft}.

\section*{Appendix: Evolution factors}
Here we reproduce for convenience of the reader the table of evolution factors for the unification scenarios and epochs considered in \cite{Part1}.
\begin{table}[h]
\hspace*{-1.5cm}
\begin{tabular}{|l|cccccc|}
\hline
\ \ Epoch        & 1    & 2     & 3    & 4    & 5      & 6  \\
\ \ \ \ \ $z_n$  & 0.14 &  0.53 &  1.6 &  3.8 & $10^3$ & $10^{10}$ \\
\hline
Scenario & $l_1\times 10^6$ & $l_2\times 10^6$ & $l_3\times 10^5$ & $l_4\times10^5$ &
	$l_5\times 10^4$ & $l_6\times 10^3$ \\

\hline
$\alpha$ only 
  & $-0.01 \pm 0.06$ & $-1.1 \pm 1.0 $  & $-0.26 \pm 0.10$ & $-0.85 \pm 0.37$&	
	$-150 \pm 350$ & $5 \pm 34$       \\
2 & $-0.1 \pm 0.1$   & $0.04 \pm 0.03$  & $-0.15 \pm 0.08$ & $0.10 \pm 0.03$ & 
	$0.9 \pm 14$   & $-0.37 \pm 0.20$ \\
3 & $ 4.1  \pm 4.8$  & $-1.5 \pm 1.2$   & $0.42 \pm 3.3$  & $-3.6 \pm 0.9$   & 
	$69 \pm 920$   & $14 \pm 8$       \\
4 & $ 3.9   \pm 8.5$ & $-3.4 \pm 2.7$   & $-8.4 \pm 5.1$  & $-8.7 \pm 2.1$   & 
	$31 \pm 450$   & $33 \pm 21$      \\
5, & $-0.02 \pm 0.18$ & $-0.24 \pm 0.18$ & $-0.25 \pm 0.10$ & $-0.61 \pm 0.13$ & 
	$0.6 \pm 8.6$  & $1.7 \pm 1.1$    \\
($\tilde{\gamma} = 42$) &&&&&& [$0.4 \pm 1.0$] \\
6, & $-0.02 \pm 0.12$ & $-0.10 \pm 0.07$ & $-0.17 \pm 0.07$ & $-0.44 \pm 0.10$ & 
	$0.3 \pm 5.0$  & $1.2 \pm 0.6$    \\
($\tilde{\gamma} = 70$) &&&&&& [$0.3 \pm 0.6$] \\
6, & $-0.12 \pm 0.18$ & $0.04 \pm 0.12$  & $-0.30 \pm 0.11$ & $0.29 \pm 0.08$  & 
	$0.7 \pm 10$   & $-1 \pm 0.3$     \\
($\tilde{\gamma}= 25$) &&&&[$-0.43 \pm 0.28$]&&\\
\hline
\end{tabular}
\caption{Redshifts and evolution factors for each epoch, for each scenario. In the first row the values of $l_n$ give the fractional variation of $\alpha$; in Scenarios 2, 5 and 6 that of $\alpha_X$; and in 3 and 4 that of $\vev{\phi}/M_X$. Values in brackets give, for BBN ($l_6$), the evolution factors neglecting \lise; or for $l_4$, the evolution factor with the Reinhold values \cite{Reinhold:2006zn} substituted by the Wendt value \cite{Wendt08}.}
\label{tab:multiplierFactors}
\end{table}

\end{document}